\documentclass[letterpaper,conference]{IEEEtran}

\IEEEoverridecommandlockouts
\def\BibTeX{{\rm B\kern-.05em{\sc i\kern-.025em b}\kern-.08em
		T\kern-.1667em\lower.7ex\hbox{E}\kern-.125emX}}

\usepackage{amsmath,amsfonts}
\usepackage{algorithmic}
\usepackage{algorithm}
\usepackage{array}
\usepackage[caption=false,font=normalsize,labelfont=sf,textfont=sf]{subfig}
\usepackage{textcomp}
\usepackage{stfloats}
\usepackage{url}
\usepackage{verbatim}
\usepackage{cite}
\usepackage{pgfplots}
\usepackage{caption}
\usepackage{graphicx,import}
\usepackage{booktabs, makecell, tabularx}
\usepackage{balance}
\newcommand\copyrighttext{%
	\footnotesize Presented at 2024 European Control Conf. DOI: \href{https://doi.org/10.23919/ECC64448.2024.10590814}{10.23919/ECC64448.2024.10590814}. \copyright 2024 IEEE.  Personal use of this material is permitted. Permission from IEEE must be obtained for all other uses, in any current or future media, including reprinting/republishing this material for advertising or promotional purposes, creating new collective works, for resale or redistribution to servers or lists, or reuse of any copyrighted component of this work in other works.}

\newcommand\copyrightnotice{%
	\begin{tikzpicture}[remember picture,overlay]
		\node[anchor=north,yshift=-5mm] at (current page.north) {\fbox{\parbox{\dimexpr\textwidth-\fboxsep-\fboxrule\relax}{\copyrighttext}}};
	\end{tikzpicture}%
}

\usepackage{graphics} 
\usepackage{epsfig} 
\usepackage{amssymb}  
\usepackage{tablefootnote}
\usepackage[pdfpagelabels,  
bookmarks,       
pdftex
]{hyperref}
\hypersetup{
	pdfborder   = 0 0 0,
	plainpages  = false,
	bookmarksnumbered = true,
}
\usepackage[nolist]{acronym}
\usepackage{balance}

\begin{acronym}
	\acro{rnn}[RNN]{Recurrent Neural Network}
	\acroplural{rnn}[RNN]{Recurrent Neural Networks}
	
	\acro{narx}[NARX-NN]{Nonlinear Autoregressive Exogenous Neural Network}
	\acroplural{narx}[NARX-NN]{Nonlinear Autoregressive Exogenous Neural Networks}
	
	\acro{gru}[GRU]{Gated Recurrent Unit}
	\acroplural{gru}[GRU]{Gated Recurrent Units}
	
	\acro{lstm}[LSTM]{Long Short-Term Memory}
	
	\acro{ann}[ANN]{Artificial Neural Network}	
	\acroplural{ann}[ANN]{Artificial Neural Networks}
	
	\acro{ffnn}[FFNN]{Feedforward Neural Network}
	\acroplural{ffnn}[FFNN]{Feedforward Neural Networks}
	
	\acro{kf}[KF]{Kalman Filter}	
	\acro{ekf}[EKF]{Extended Kalman Filter}
	\acro{ukf}[NEKF]{Neural Extended Kalman Filter}
	\acro{ukf}[UKF]{Unscented Kalman Filter}
	\acro{hekf}[H-EKF]{hybrid Extended Kalman Filter}
	
	\acro{knn}[$K$NN]{$K$-Nearest-Neighbors}
	
	\acro{mftm}[MFTM]{Magic Formula Tire Model}
	
	\acro{cog}[COG]{Center Of Gravity}
	\acro{ebs}[EBS]{Electronic Braking System}
	\acro{lti}[LTI]{Linear Time-Invariant}
	\acro{pso}[PSO]{Particle Swarm Optimization}
	\acro{nmse}[NMSE]{Normalized Mean Squared Error}
	\acro{mse}[MSE]{Mean Squared Error}
	\acro{rmse}[RMSE]{Root Mean Squared Error}
	\acro{ml}[ML]{Machine Learning}
	
	\acro{rpe}[RPE]{Recursive Predictive Error}
	
	\acro{adas}[ADAS]{Advanced Driver Assistance Systems}
	\acroplural{adas}[ADAS]{Driver Assistance Systems}
\end{acronym}

\newcommand{\del}{\partial}
\newcommand{\bi}[1]{\boldsymbol{#1}}
\newcommand{\ur}[1]{\mathrm{#1}}
\newcommand{\cali}[1]{\mathcal{#1}}

\newcommand\norm[1]{\left\lVert#1\right\rVert}
\newcommand{\km}{{k \scalebox{0.75}[0.75]{\( - \)} 1}}

\newcommand{\xcog}{l_{2,{\ur{COG}}}}

\newcommand*{\tr}{^{\top}}

\newcommand*{\Zeros}{0}
\newcommand*{\ANN}{\mathrm{ANN}}

\mathchardef\mhyphen="2D
\newcommand*{\ie}{i.\,e.,\,}
\newcommand*{\eg}{e.\,g.,\,}

\definecolor{imesgruen}{RGB}{200,	211,	23}
\definecolor{imesblau}{RGB}{0,	80,	155}
\definecolor{imesorange}{RGB}{231,	123,	41}

\begin{document}
	

\title{Reliable State Estimation in a Truck-Semitrailer Combination using an Artificial Neural Network-Aided Extended Kalman Filter}

\author{Jan-Hendrik~Ewering, Zygimantas~Ziaukas, Simon~F.~G.~Ehlers, and Thomas~Seel
	\thanks{The Authors are with the Institute of Mechatronic Systems, Leibniz Universit\"at Hannover, 30167~Hanover, Germany (e-mail: \href{mailto:jan-hendrik.ewering@imes.uni-hannover.de}{jan-hendrik.ewering@imes.uni-hannover.de}).}%
}

\maketitle\copyrightnotice\vspace{-9.5pt}

\begin{abstract}
Advanced driver assistance systems are critically dependent on reliable and accurate information regarding a vehicles' driving state. 
For estimation of unknown quantities, model-based and learning-based methods exist, but both suffer from individual limitations. 
On the one hand, model-based estimation performance is often limited by the models' accuracy. 
On the other hand, learning-based estimators usually do not perform well in ``unknown'' conditions (bad generalization), which is particularly critical for semitrailers as their payload changes significantly in operation. 
To the best of the authors' knowledge, this work is the first to analyze the capability of state-of-the-art estimators for semitrailers to generalize across ``unknown'' loading states. 
Moreover, a novel \ac{hekf} that takes advantage of accurate \ac{ann} estimates while preserving reliable generalization capability is presented. It estimates the articulation angle between truck and semitrailer, lateral tire forces and the truck steering angle utilizing sensor data of a standard semitrailer only. 
An experimental comparison based on a full-scale truck-semitrailer combination indicates the superiority of the \ac{hekf} compared to a state-of-the-art \acl{ekf} and an \ac{ann} estimator.

\end{abstract}

\acresetall
\section{Introduction}
\IEEEPARstart{R}{eliable} and accurate information about a vehicles' driving state is crucial for safe \ac{adas} in automotive applications. In this light, model-based and learning-based estimation algorithms are employed to infer unknown quantities.
To condition these algorithms for a specific system, an initial training process (learning-based estimation) or a model identification procedure (model-based estimation) based on a representative training data set is usually conducted. 
However, obtaining an extensive training data set that covers the whole operation region is difficult and sometimes impossible. 
Reasons can be that exhaustive driving tests are too expensive, not all use cases are reproducible in laboratory conditions, or that unforeseen situations may occur during the life cycle of a product (\eg changing system parameters or environmental conditions). 
In this context, reliable estimation in ``unknown'' situations (\ie when the current inputs and measurements differ from the training data distribution) is critical. In the following, we will refer to the capability to generalize in these situations as ``out-of-distribution'' generalization. 
In contrast, we use the term ``in-distribution'' evaluation, if estimation methods are employed in situations, when the current inputs and measurements are similar to the training data set.\\
Good generalization capability of estimation methods ``out-of-distribution'' is especially relevant for commercial vehicles, \eg semitrailers, as (i) the payload varies significantly during operation and (ii) driving tests are associated with great effort and expensive. Existing works for estimation in truck-semitrailer combinations can be categorized in model-based, learning-based and combined model- and learning-based (hybrid) approaches.
\begin{figure}[t]
	\centering
	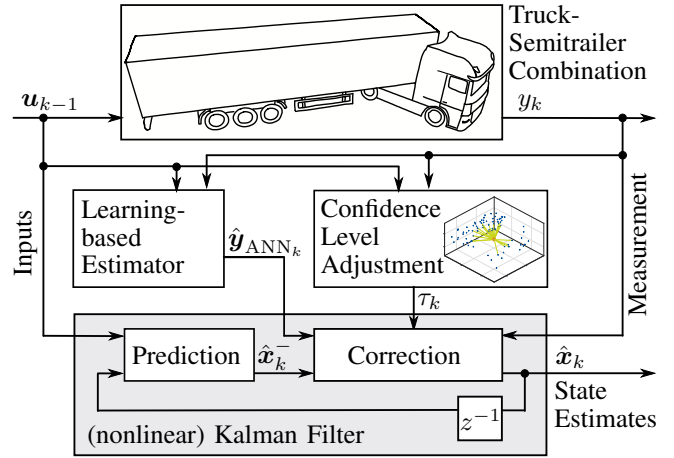
	\caption{Hybrid estimation scheme for reliable combination of model-based and learning-based estimators, using the confidence $\tau_k$ in the learning-based method. The confidence $\tau_k$ in the learning-based estimate $\hat{\bi{y}}_{\ur{ANN}_k}$ is determined by the similarity between the current operation point and the training data set. $z^{-1}$ denotes a delay by one time instant.}
	\label{fig:hyb_ekf}
\end{figure}\\
For \textbf{model-based estimation} in truck-semitrailer combinations, most previous works rely on nonlinear Kalman filtering. The authors of \cite{AhmadiJeyed.2019} use an \ac{ekf} \cite{Sarkka.2023} for estimation of lateral dynamic quantities, validating their method by means of a detailed simulation with a single loading state. In \cite{Ziaukas.2019}, an \ac{ekf} is applied for estimation in a real-world truck-semitrailer combination. The authors employed a single-track model of the lateral dynamics and identified it for one particular payload. A more elaborate approach based on a two-track model and an \ac{ukf} \cite{Sarkka.2023} is presented in \cite{Ehlers.2022}. Here, three distinct loading states are used for model identification and the experimental results are carried out for in-distribution scenarios. 
Although model-based methods usually have good generalization capability out-of-distribution, existing works do not exploit this advantage for estimation in truck-semitrailer systems.\\
On the other hand, \textbf{learning-based} methods can yield excellent estimation results. 
In the context of truck-semitrailer systems, a \ac{rnn} is used in \cite{Jahn.2020} for estimation of the articulation angle and the steering angle. Experimental data of a single loading state is considered for training and evaluation of the \ac{rnn}. While the learning-based estimator in \cite{Jahn.2020} can outperform an \ac{ekf} in-distribution, it remains an open question, whether its generalization capability to other conditions (\eg other payloads) is sufficient for reliable operation, as pointed out by the authors.\\ 
\textbf{Hybrid model-based and learning-based} estimation schemes attempt to combine the advantages of both approaches. In \cite{Graber.2019}, the side-slip angle in a car is estimated by an \ac{ann} whose input features are generated from partial physical model knowledge. The approach yields better estimation accuracy than a solely learning-based estimator, but has difficulties to generalize robustly to different road-tire friction conditions.\\
Other works augment (nonlinear) derivates of the \acl{kf} with learning-based methods \cite{Kim.2022,Revach.Jun.2021,Korayem.2020}. An overview can be found in \cite{Kim.2022}. A particularly interesting hybrid approach is the confidence-based hybrid estimation scheme presented in \cite{Sieberg.2022}, in which sensor information is preprocessed by an \ac{ann} to obtain estimates of a car's roll angle. A confidence measure in the \ac{ann} estimate is computed based on the similarity between current input data and training data. Considering this confidence, the ``soft measurements'' are used in a \ac{ukf} \cite{Sieberg.2022b,Sieberg.2022}.
In the context of truck-semitrailer systems, various hybrid estimation architectures combining \ac{ann} and \ac{ukf} are designed and evaluated in \cite{Ehlers.2023b} based on an extensive real-world data set with different semitrailer loading states. In the paper, it is found that the best scheme for this application case and data set is an \ac{ann} that corrects the estimates of a \ac{ukf}.
However, as pointed out in \cite{Ehlers.2023b}, the robustness to changed conditions (\eg different payloads) is limited and requires further investigation.\\
In the literature, few works analyze and improve the reliability of estimation methods out-of-distribution. For truck-semitrailer combinations, no work is concerned with generalization capability regarding payload variation, and no existing methods allow for well-generalizing learning-based or hybrid estimation across various unknown payloads. 
{In this light, the current work is concerned with the generalization capability of estimators out-of-distribution. The research is conducted for a truck-semitrailer system, as the variation of payload is a good example for out-of-distribution problems, but the algorithms may be utilized in other applications as well. In the current problem setup, we aim to estimate the articulation angle between truck and semitrailer, lateral tire forces and the truck steering angle utilizing sensor data of a standard semitrailer only.}\\
In particular, the contributions of this work are: \textbf{(i)} an extensive experimental analysis of state-of-the-art estimation methods in- and out-of-distribution using a real-world data set, \textbf{(ii)} a novel \ac{hekf} for truck-semitrailer combinations (see Fig.~\ref{fig:hyb_ekf}), enabling accurate estimation in-distribution and ensuring reliable generalization out-of-distribution, \textbf{(iii)} extending the confidence-based estimation framework \cite{Sieberg.2022} regarding multidimensional \ac{ann} estimates and \textbf{(iv)} a novel \ac{knn}-based approach for determination of the confidence $\tau_k$.\\
The paper is structured as follows. Sec.~\ref{sec:modeling} is concerned with modeling of the truck-semitrailer system. In Sec.~\ref{sec:estimation}, existing estimation methods are revisited, and the \ac{hekf} estimation scheme is discussed in detail. An extensive experimental analysis based on real-world data in- and out-of-distribution is presented in Sec.~\ref{sec:experimental_results}. Last, a conclusion and an outlook are given in Sec.~\ref{sec:conclusion}. 

\section{Modeling a Truck-Semitrailer Combination}
\label{sec:modeling}
The model-based \ac{ekf} in this work relies on a nonlinear single-track model of a truck-semitrailer combination's lateral dynamics. The original model is derived and presented in detail in \cite{Ziaukas.2019}. For better generalization capability across various loading states, some enhancements are introduced in this contribution. To retain a concise presentation of the findings, we focus on the adjusted model parts in the following.\\
The system model is derived from rigid body dynamics and contains two subparts, the truck and the semitrailer, which are linked by a constraint equation and denoted by subscripts $i=1$ for the truck and $i=2$ for the semitrailer. From the governing equations in lateral direction, expressions for the lateral velocities ${v}_{y_i}$ and the yaw rates $\dot{\psi}_i$ can be obtained \cite{Ziaukas.2019}. The articulation angle between the truck and the semitrailer obeys $\dot{\theta} = \dot{\psi}_2  - \dot{\psi}_1$. Besides, the model is influenced by the steering angle of the truck's first axle $\delta_1$ and the longitudinal velocity $v_{x_2}$.\\
Moreover, the lateral tire forces $F_{y_{ij}}$ need to be considered in the lateral dynamics model. In this contribution, the basic tire model used in \cite{Ziaukas.2019} is replaced by the \ac{mftm} \cite{Pacejka.2012} to increase accuracy. Thus, the lateral tire forces are modeled by the differential equations
\begin{align} \label{eq:mtmf}
	& F_{y_{ij}} + \frac{l_{ij}}{v_{x_i}} \dot{F}_{y_{ij}} = \mu_{\ur{max}} F_{z_{ij}} \sin \left( C_{ij} \arctan \left( B_{ij} \tan \left( \alpha_{ij} \right) \right) \right), \\
	& B_{ij} = {c_{1_{ij}} \sin \left( 2 \arctan \left( F_{z_{ij}} / c_{2_{ij}} \right)\right)}/\left({C_{ij} D_{ij}}\right),
\end{align}
where all quantities with two scalar subscripts $\square_{ij}$ refer to the $j$-th axle in rigid body $i$, and the tire model parameters $C_{ij}$, $D_{ij}$, $c_{1_{ij}}$, $c_{2_{ij}}$ and $l_{ij}$ can be obtained from identification. $\alpha_{ij}$ are the slip angles, and $\mu_{\ur{max}}$ is the maximum road friction coefficient.\\
As can be seen in \eqref{eq:mtmf}, the lateral tire forces are dependent on the vertical axle forces $F_{z_{ij}}$, which are assumed fixed in the original model \cite{Ziaukas.2019}. In contrast, different loading states of the semitrailer, and thus varying vertical axle forces $F_{z_{2j}}$, are considered in this contribution. 
The vertical axle forces are modeled by an equal force distribution 
\begin{equation} \label{eq:vert_forces}
	F_{z_{2j}} =  {F_{z_2}}/{3},
\end{equation}
based on the overall axle load $F_{z_2}$, which is determined from serial air suspension pressure measurements. 
Besides, a static vertical model of the semitrailer is used to represent the relationship between $F_{z_2}$, the semitrailer's mass $m_2$, and its center of gravity along the longitudinal axis $\xcog$ as
\begin{equation} \label{eq:vert_forces2}
	m_2 =  {l_{\ur{Agg}} F_{z_2}}/{\xcog},
\end{equation}
where $l_{\ur{Agg}}$ is the longitudinal distance from the connection point between truck and trailer (king pin) to the center of the running gear. 
The resulting state space model features the state vector $\bi{x}_{\ur{SSM}}~\in~\mathbb{R}^{n_{\ur{SSM}}}$ with
\begin{equation}
	\bi{x}_{\ur{SSM}}\tr=\left[v_{y_1}, \dot{\psi}_{1}, v_{y_2}, \dot{\psi}_{2}, \theta, F_{y_{11}}, F_{y_{12}}, F_{y_{21}}, F_{y_{22}}, F_{y_{23}}\right].
\end{equation}
In total, the physical model exhibits $15$ parameters to be identified, most of which are tire parameters from \eqref{eq:mtmf}. 
The parameter identification is performed using a \ac{pso} \cite{Kennedy.1995} following the lines of \cite{Ziaukas.2019, Ehlers.2022}. 
The associated cost function is the \ac{nmse} between model simulation results and available ground truth state measurements in the test vehicle.

\section{Estimation Methods for a Truck-Semitrailer Combination}
\label{sec:estimation}  
The traditional methods \ac{ekf} (model-based) and \ac{ann} (learning-based) form a basis for the presented hybrid \ac{ekf}. Thus, we briefly revisit both estimation schemes in the context of the current truck-semitrailer system in Sec.~\ref{sec:model_driven_estimation} and \ref{sec:data_driven_estimation}, taking into account previous results \cite{Ziaukas.2019,Jahn.2020}. Thereafter, the hybrid \ac{ekf} is introduced and explained in detail.\\
As stated in the introduction and following the lines of previous publications \cite{Ziaukas.2019,Jahn.2020,Ehlers.2023b,Ehlers.2022}, the problem setting is to estimate the articulation angle between truck and semitrailer $\theta$, the lateral tire forces $F_{y_{2j}}$ and the truck steering angle $\delta_1$ utilizing sensor data of a standard semitrailer only. 

\subsection{Model-based Estimation}
\label{sec:model_driven_estimation}
For reliable estimation based on standard semitrailer sensors only, information regarding the current loading state (\ie information about the semitrailer's vertical axle forces $F_{z_{2j}}$, its mass $m_2$, and the center of gravity $\xcog$) is crucial. In this light, conventional semitrailers are equipped with an air suspension and an air braking system, both controlled by onboard computing units. The corresponding pressure sensor information is used in this contribution to determine the current summed axle forces of the semitrailer $F_{z_2}$. Taking into account \eqref{eq:vert_forces2}, the mass $m_2$ can be calculated based on an assumption about the center of gravity position $\xcog$, which is obtained by estimation in this contribution. 
It is worth mentioning that the estimated \ac{cog} and the resulting semitrailer mass $m_2$ are only auxiliary quantities which are not necessarily physically meaningful. However, estimating $\xcog$ adds a degree of freedom, which can account for changed loading conditions and increase overall estimation accuracy. 
Moreover, the steering angle input $\delta_{1}$ needs to be estimated, as it is not available from sensor measurements in a standard semitrailer.\\
An ad-hoc approach for practical estimation of unknown parameters and inputs is to model the unknown quantities with a random walk and to augment the state vector accordingly \cite{Alberding.2013,Sarkka.2023}. Thus, the extended state vector $\bi{x}~\in~\mathbb{R}^{n}$ and the input vector $\bi{u} ~\in~\mathbb{R}^{m}$ for model-based estimation read
\begin{align}
	\bi{x}\tr &= \begin{bmatrix}
		\bi{x}_{\ur{SSM}}\tr & \delta_1 & \xcog
	\end{bmatrix},\\
	\bi{u}\tr &= \begin{bmatrix}
	v_{x_2} & F_{z_2}
\end{bmatrix},
\end{align}
respectively. For purely model-based estimation, a single measurement $y = {y}_{\mathrm{EKF}} = \dot{\psi}_2$ is available\footnote{In operation, the yaw rate $\dot{\psi}_2$ is computed from wheel speed and wheel speed difference of left and right wheels at the semitrailer's middle axle.}. Because the subsequent filtering equations are relevant for the hybrid \ac{ekf} with an augmented measurement vector $\bi{y}_{\mathrm{HEKF}}$ as well (see Sec.~\ref{sec:combined_estimation}), the general multidimensional equations are introduced in the following.\\
Using the original derivation in \cite{Ziaukas.2019} together with the presented modeling choices, a discrete-time state space representation
\begin{subequations} \label{eq:model}
\begin{align} 
	\bi{x}_{k}  & =  \bi{f}_{\ur{d}} ( \bi{x}_{\km}, \bi{u}_{\km}) + \bi{v}_{\km}, \;  
	& \bi{v}_{\km} \sim \mathcal{N} \left( \Zeros,   \bi{Q}	\right), \label{eq:modela}\\ 
	\bi{y}_{k}  & =  \bi{g}_{\ur{d}} (\bi{x}_{k}) + \bi{w}_{k},  
	& \bi{w}_{k} \sim \mathcal{N} \left( \Zeros,  \bi{R} \right), \label{eq:modelb}
\end{align}
\end{subequations}
of the truck-semitrailer system is obtained, where $\bi{f}_{\ur{d}} : \mathbb{R}^{n} \times \mathbb{R}^{m} \rightarrow  \mathbb{R}^n$ is the state transition function, $\bi{g}_{\ur{d}} : \mathbb{R}^{n} \rightarrow  \mathbb{R}^{p}$ is the output function and $\bi{y}_k~\in~\mathbb{R}^{p}$ is the vector of system outputs at time step $k$. $\bi{v}_k$ and $\bi{w}_k$ are additive process and observation noise which are assumed to be drawn from zero mean, independent Gaussian distributions with covariance matrices $\bi{Q}$ and $\bi{R}$, respectively. In this contribution, the covariance matrix $\bi{R}$ is determined based on measurements and the state error covariance matrix is tuned heuristically.\\
A standard approach for recursive Bayesian state estimation in real-world applications is Kalman filtering. 
In linear systems with Gaussian noise, the method provides a closed-form solution to the optimal filtering equations \cite{Sarkka.2023}. 
Despite its simplicity, Kalman filtering provides remarkable accuracy in a wide range of applications. 
In a prediction step, \textit{a priori} information about inputs and states is used together with the system model to predict the system state and its error covariance in the next time instant. In a subsequent correction step, evidence about the true system behavior (\ie measurements) is incorporated using Bayes' rule to determine an \textit{a posteriori} state and error covariance estimate \cite{Sarkka.2023}. A recursive algorithm is obtained by employing the estimates as the prior for the next time instant.\\
To deal with nonlinear systems, various derivates of the Kalman filter, such as the \ac{ekf} or the \ac{ukf}, are available \cite{Sarkka.2023}. 
In this work, an \ac{ekf} is employed, given that previous results have shown its applicability in the current nonlinear system \cite{Ziaukas.2019}. The \ac{ekf} accounts for nonlinearities by linearizing the system model in each time step $k$ at the current operation point. The discrete time \ac{ekf} equations in the presence of additive noise are
\begin{subequations} \label{eq:pred}
\begin{align} 
	\hat{ \bi{x} }_{k}^{-} &=
	\bi{f}_{\ur{d}} (\hat{ \bi{x} }_{\km},\bi{u}_{\km}), \label{eq:preda}\\
	\bi{P}_{k}^{-} & = \bi{A}_{k} \bi{P}_{\km} \bi{A}_{k}\tr + \bi{Q}, \label{eq:predb}
\end{align}
\end{subequations}
for the prediction step and
\begin{subequations} \label{eq:correction}
\begin{align} 
	\bi{K}_{k} & =  \bi{P}_{k}^{-} \bi{C}_k\tr \left( \bi{C}_k \bi{P}_{k}^{-}  \bi{C}_k\tr + \bi{R} \right)^{-1}\ur{,} \label{eq:correctiona}\\
	\hat{ \bi{x} }_{k} & = \hat{ \bi{x} }_{k}^{-} + \bi{K}_{k} \left( \bi{y}_{k} - \bi{C}_k \hat{ \bi{x} }_{k}^{-} \right)\ur{,} \label{eq:correctionb}\\
	\bi{P}_{k} & = \left( \mathbf{I} - \bi{K}_{k} \bi{C}_k \right) \bi{P}_{k}^{-}, \label{eq:correctionc}
\end{align}
\end{subequations}
for the correction step \cite{Sarkka.2023}. The superscripts $\square^-$ denote quantities computed in the prediction step. $\hat{ \bi{x} }_{k} \in \mathbb{R}^n $ is the estimated state vector at time step $k$. $\bi{P}_k$ is the estimated error covariance matrix, and the Kalman gain is denoted $\bi{K}_k$. The identity matrix of suitable dimension is $\mathbf{I}$.\\
The momentary state transition matrix $\bi{A}_{k}$ and the state decoupling matrix $\bi{C}_{k}$ are obtained by linearization at the current operation point from
\begin{equation} \label{eq:Jacobians}
	\bi{A}_k = \left. \frac{\del \bi{f}_{\ur{d}} }{\del \bi{x} } \right|_{\hat{\bi{x}}_{ \km}, \bi{u}_{\km}}, \qquad 
	\bi{C}_k = \left. \frac{\del \bi{g}_{\ur{d}} }{\del \bi{x} } \right|_{\hat{\bi{x}}_{k}^{-}}.	
\end{equation}

\subsection{Learning-based Estimation}
\label{sec:data_driven_estimation}
Apart from purely model-based algorithms, estimation can be conducted using learning-based approaches as well. The learning-based approach in this contribution relies on an \acp{rnn}. These \ac{ann} structures are widely used for estimation in dynamical systems due to their capability to learn subtle effects and patterns in time series data. 
However, \ac{rnn} require special attention regarding their training, which can become unstable \cite{Du.2014}. Modern \ac{rnn} structures, like \ac{lstm} cells and \acp{gru}, have been designed to mitigate these difficulties \cite{Calin.2020}.\\
In this contribution, we use a specific \ac{rnn} structure, a \ac{narx}, as this structure has been successfully utilized for state estimation in the truck-semitrailer system in a previous work \cite{Jahn.2020}. 
In \ac{narx}, recurrence is induced by feeding the \ac{ann} outputs of the last time instant back to the input in the current time step, resulting in a recursion loop. For training of \ac{narx}, it is convenient to apply two phases, first in open loop setting and second in closed loop configuration \cite{Jahn.2020}.\\
In the current setting, the inputs $\bi{u}_{\ANN}$ and the outputs $\hat{\bi{y}}_{\ANN}$ (\ie the learning-based estimates) of the \ac{narx} are 
\begin{align}
	\bi{u}_{\ANN} &= \begin{bmatrix}
		v_{x_2} & F_{z_2} & \dot{\psi}_2
	\end{bmatrix}\tr, \\
	\hat{\bi{y}}_{\ANN} &= \begin{bmatrix}
		\dot{\psi_{1}} & \theta & F_{y_{21}} & F_{y_{23}} & \delta_{1}
	\end{bmatrix}\tr,
\end{align}
respectively. 
To capture more information contained in consecutive time steps, two input and two feedback delays are used, which is a typical modification for sequence estimation with \ac{ann} \cite{Du.2014}.\\ 
The training procedure starts with one epoch in open-loop setting and continues with a maximum of $1000$ epochs in closed loop, following the lines of \cite{Jahn.2020}. The configuration of the \acp{ann} is found by a grid search between one and three layers and up to 30 neurons in total. Tangent hyperbolicus activation functions are chosen for all hidden layers, and the output layer has a linear activation. The employed optimization algorithm is Bayesian regularization backpropagation.\\
A separate \ac{ann} is trained for each estimated variable to keep the complexity of the networks and the training process low and to reduce possible negative cross-effects. 
\begin{figure*}[t]
	\centering
	\parbox{0.95\textwidth}{\includegraphics[width=0.95\textwidth]{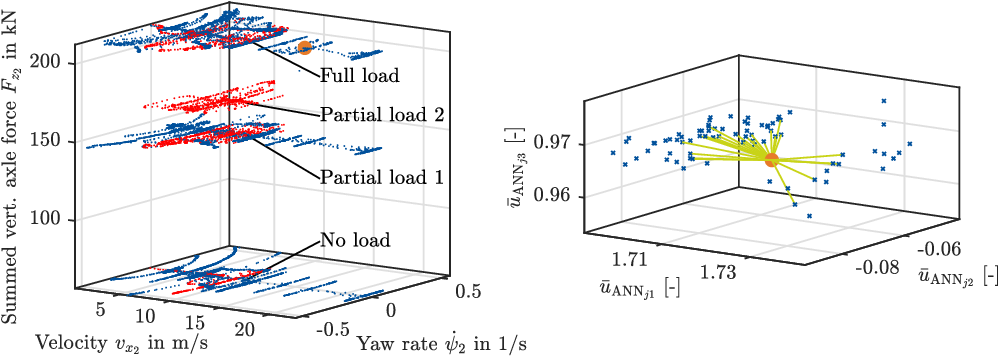}}
	\caption{\ac{ann} training inputs $\{\bi{u}_{\ANN_j}^{\ur{tr}} \}_{j=1}^{N_{\ur{tr}}}$ (blue) and evaluation inputs (red, orange)  
	from raw data set (left) and excerpt of standardized data set (right) around new evaluation input $\bar{\bi{u}}_{\ANN_k}$ (orange) with \ac{knn}-search for $K = 25$. The evaluation data of loading states ``full load'', ``partial load 1'', and ``no load'' is in-distribution regarding the training data. The evaluation data of loading state ``partial load 2'' is out-of-distribution regarding the training data.}
	\label{fig:knn_input_space}
\end{figure*}
\subsection{Hybrid Estimation}
\label{sec:combined_estimation}
A combination of model- and learning-based methods for state estimation can be advantageous to increase estimation accuracy while preserving reliability and physical interpretability \cite{Sieberg.2022b,Sieberg.2022}. 
Therefore, we employ \ac{narx} to estimate $\hat{\bi{y}}_{\ANN}$ and consider the estimates as ``soft measurements''. The original measurement vector $y$ is extended such that additional information based on the \ac{narx} is provided in the correction step of the \ac{hekf}. Thus, the \ac{hekf} measurement vector for estimation in the truck-semitrailer system is
\begin{align}
	\bi{y}_{\ur{HEKF}} &= \begin{bmatrix}
		y & \hat{\bi{y}}_{\ANN}\tr
	\end{bmatrix}\tr.
\end{align}
Based on \cite{Sieberg.2022b,Sieberg.2022}, the confidence $\tau_k$ in the ``soft measurements'' is used to dynamically adjust their impact on the current \ac{hekf} estimate. 
For instance, the confidence in a ``soft measurement'' sample $\hat{\bi{y}}_{\ANN_k}$ is assumed high if the input data $\bi{u}_{\ANN_k}$ in time instance $k$ is similar to the training data. In turn, if the current input data is not similar to the training data, the confidence $\tau_k$ is low. An overview of the \ac{hekf} can be seen in Fig. \ref{fig:hyb_ekf}.\\
To calculate the confidence, a histogram approach is proposed in \cite{Sieberg.2022b,Sieberg.2022}. Following this approach, the input training data $ \{ \bi{u}_{\ANN_j}^{\ur{tr}} \} _{j=1}^{N_{\ur{tr}}} $, $\bi{u}_{\ANN_j}^{\ur{tr}} \in \mathbb{R}^{m_{\ANN}}$, is binned to $(n_{\ur{grid}})^{m_{\ANN}}$ subspaces, and the resulting bin count is used to reason about the confidence in the \ac{ann} at different points in the input space. Please note, the superscript $\square^{\ur{tr}}$ is used for quantities that are employed in \ac{ann} training. Depending on $\tau_k$, the matrices $\bi{Q}_{k,\ur{UKF}}$ and $\bi{R}_{k,\ur{UKF}}$ of an \ac{ukf} are adjusted in \cite{Sieberg.2022} to incorporate more or less knowledge from the ``soft measurements'', respectively.\\
However, the binning approach employs distinct boundaries in the input data space which leads to discontinuous confidence jumps, \ie data points opposite to each other just at the border between two bins may vary significantly in confidence regardless of their similarity. Preferably, the confidence should vary smoothly. To achieve this, a \ac{knn}-based approach is exploited here (see Fig. \ref{fig:knn_input_space}).\\
\ac{knn} is a supervised machine learning technique for classification of data. In the method, a labeled training data set $\cali{D}_\ur{d} = \{\bi{s}_j^\ur{d}, l_j^\ur{d}\}_{j=1}^{N_\ur{d}}$ with sample points $\bi{s}_j^\ur{d}$ and labels $l_j^\ur{d}$ is compared with an unlabeled query point $\bi{s}_k$ which should be classified. The query point's $K$ nearest neighbors in $\cali{D}_\ur{d}$ are found, and the most probable label for $\bi{s}_k$ is inferred from the training data labels \cite{Cherkassky.2007}. In the context of the estimation task, the training data set 
\begin{align}\label{eq:data_set}
	\cali{D}_{\ur{tr}}~&=~\left\{\bi{u}_{\ANN_j}^{\ur{tr}}, \bi{y}_{\ANN_j}^{\ur{tr}}\right\}_{j=1}^{N_{\ur{tr}}},
\end{align}
is compared to the current \ac{ann} input in operation $\bi{u}_{\ANN_k}$. The $K$ nearest neighbors in the standardized data set are found, and the corresponding mean squared Euclidean distance $d_{k}$ is computed according to
\begin{equation} \label{eq:distance}
	d_{k}  = \frac{1}{K} \sum_{j=1}^{K} \norm{\bar{\bi{u}}_{\ANN_j}^{\ur{tr}} - \bar{\bi{u}}_{\ANN_k} }_2^2,
\end{equation}
where the bar symbol $\bar{\square}$ denotes a standardized quantity.  
The confidence $\tau_{k}$ in the ``soft measurement'' $\hat{\bi{y}}_{\ANN_{k}}$ is computed
\begin{equation} \label{eq:confidence}
	{\tau_{k}}  = 
	\begin{cases}
			(d_{\ur{max}} - d_{k})/d_{\ur{max}},  &\text{if}\ d_{k} \leq d_{\ur{max}},\\
			0, & \text{otherwise,}
	\end{cases}
\end{equation}
where the maximum distance $d_{\ur{max}}$ is a threshold above which no confidence in the ``soft measurements'' is present.\\
During operation of the \ac{hekf}, the current confidence $\tau_k$ is used to compute the measurement noise covariance matrix $\bi{R}_k$. A quadratic mapping between $\tau_k$ and $\bi{R}_0$ yields good experimental results for the current application to a truck-semitrailer system. Thus, we choose
\begin{equation} \label{eq:R_adjustment}
	\bi{R}_k =  \left( c \left( \tau_k - 1 \right)^2 +1 \right) \cdot \bi{R}_0,
\end{equation}
for the subsequent findings, where $c$ is a parameter that determines the measurement noise covariance at zero confidence.\\
For tuning, the covariance matrices of the \ac{hekf} are determined initially with all ``soft measurements'' available (\ie $\tau_k = 1 \; \forall \; k$) such that the filter is stable, and the \ac{hekf} results follow $\hat{\bi{y}}_{\ANN_k}$ closely. The resulting covariance matrices are stored as $\bi{Q}$ and $\bi{R}_0$.

\section{Experimental Results}
\label{sec:experimental_results}
The performance of the \ac{hekf} (Sec.~\ref{sec:combined_estimation}) in- and out-of-distribution is compared to state-of-the-art estimation methods for truck-semitrailer systems, \ie the purely model-based \ac{ekf} (Sec.~\ref{sec:model_driven_estimation}) and the learning-based \ac{ann} (Sec.~\ref{sec:data_driven_estimation}), using an extensive experimental data set of a truck-semitrailer combination with different loading states. The data set and the parameterization procedure are described first. 
Second, the experimental results are presented and discussed.

\subsection{Data Set and Parameterization}
For model identification and training of the \ac{ann}, extensive driving experiments are carried out with a real-world truck-semitrailer combination. The experimental vehicle is similar to the one presented in \cite{Ziaukas.2019, Jahn.2020, Ehlers.2022, Ehlers.2023b}. In particular, to measure the lateral tire forces $F_{y_{21}}$ and $F_{y_{23}}$, the first and the third axle of the experimental semitrailer have been equipped with a calibrated and verified test setup based on strain gauges. The available sensor measurements in the test vehicle and in an off-the-shelf three-axle semitrailer are summarized in Tab. \ref{tab:variables} for an overview. 
The only signals used for estimation are the semitrailers' longitudinal velocity $v_{x_2}$, yaw rate $\dot{\psi_{2}}$ and the summed vertical axle force $F_{z_2}$, which are available in a standard semitrailer. All other signals defined in Tab.~\ref{tab:variables} are used for \ac{ann} training and model identification only.
\begin{table}
	\caption{Measurements available in Test Vehicle and Standard Product (S denotes Semitrailer, T denotes Truck)}
	\label{tab:variables}
	\begin{center}
		\begin{tabular}{l l c c}
			\hline
			\hline
			Origin 	& Measured variable & Symbol & Available in\\
			& 					&  &standard product\\
			\hline
			S & longitudinal velocity		& $v_{x_2}$ 		& Yes\\
			S & yaw rate   					& $\dot{\psi}_2$ 	& Yes\\
			S & summed vertical axle force   				& $F_{z_2}$ 		& Yes\\
			S & lateral tire force			& $F_{y_{21}}$ 		& No\\
			S & lateral tire force			& $F_{y_{23}}$ 		& No\\
			S/T & articulation angle		& $\theta$ 			& No\\
			T & longitudinal velocity		& $v_{x_1}$ 		& No\\
			T & yaw rate 					& $\dot{\psi}_1$ 	& No\\
			T & steering angle 				& $\delta_{1}$ 			& No\\
			\hline
			\hline
		\end{tabular}
	\end{center}
\end{table}\\
The driving maneuvers in the training and identification data set are selected to be specifically exciting to the lateral dynamics of the truck-semitrailer combination, which includes sinusoidal, step and ramp steering angle inputs (up to $\pm \, 30 ^{\circ}$) at different velocities ($3$ to $22$ $\mathrm{m/s}$) and with different loading states. The different loading states correspond to multiple payloads with individual mass and spatial distribution.
Specifically, the training data $\cali{D}_{\ur{tr}}$ comprises three full sets of similar driving maneuvers for the loading states ``full~load'' (payload $21,600\,\ur{kg}$), ``partial load 1'' (payload $16,000\,\ur{kg}$) and ``no load'' (payload $0\,\ur{kg}$) with different spatial distributions of the payload.\\ 
For evaluation of the estimation methods, additional maneuvers covering a broad area in the operation space are available. These maneuvers are carried out while the semitrailer is loaded as for acquisition of the training data (in-distribution evaluation). A fourth maneuver is performed with a different loading state ``partial load 2'' (payload $16,000\,\ur{kg}$) featuring a changed payload distribution. With this maneuver, we evaluate the estimation performance in operational areas that are not covered in the training data $\cali{D}_{\ur{tr}}$ (see Fig.~\ref{fig:knn_input_space}, left). Thus, the generalization error and reliability in ``unknown'' scenarios can be tested (out-of-distribution evaluation).
\begin{figure*}[ht]
	\centering
	\parbox{1\textwidth}{\includegraphics[width=1\textwidth]{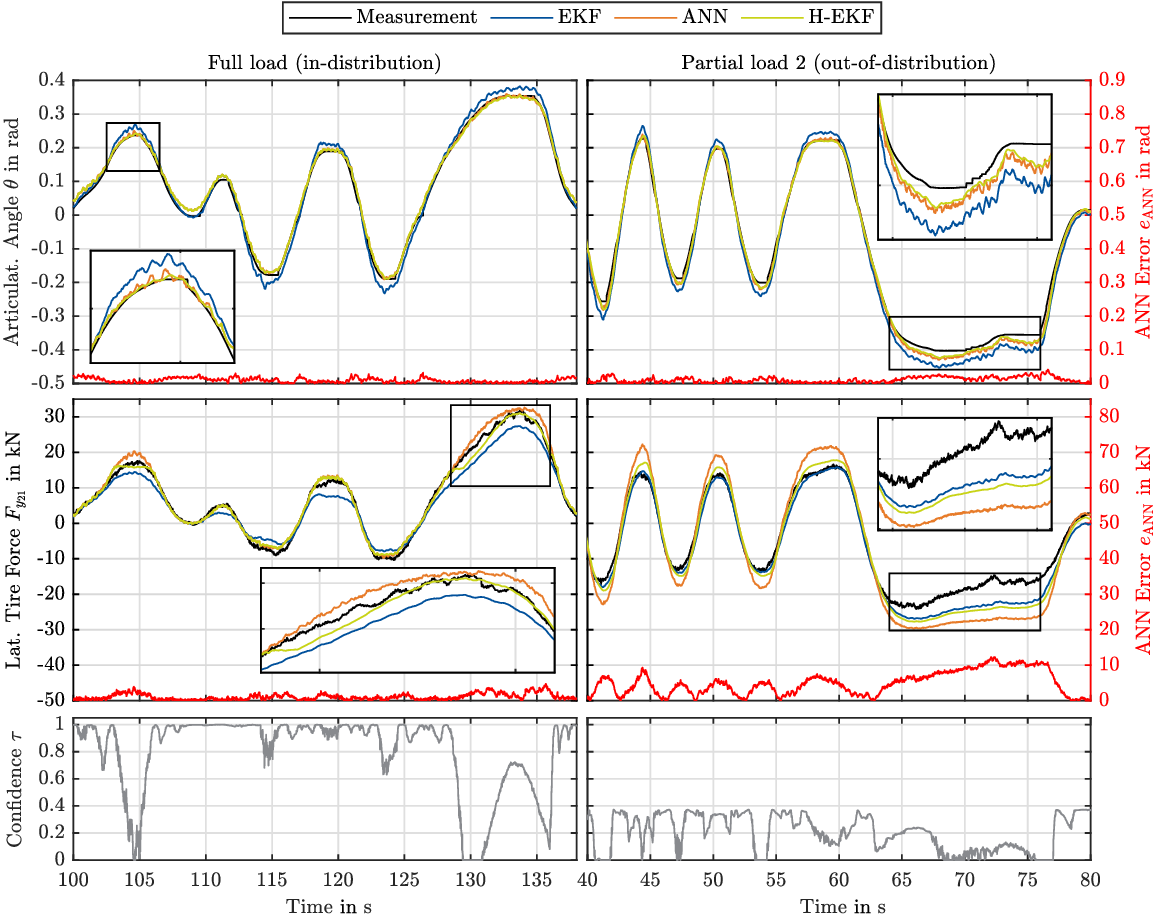}}
	\caption{Measurements and estimates of articulation angle $\theta$ (top) and lateral tire force $F_{y_{21}}$ (middle) with \ac{ann} estimation errors and current confidences (bottom). Excerpts from evaluation data of loading state ``full load'' (left) and ``partial load 2'' (right). The \ac{ann} performs well if the evaluation data is close to the training data (\ie the confidence $\tau$ is high) and shows large errors if the confidence is close to $0$. The \ac{hekf} incorporates \ac{ann} estimates and improves estimation accuracy reliably by taking the confidence into account.}
	\label{fig:exp_results_4}
\end{figure*}

\subsection{Experimental Estimation and Generalization Analysis}
For illustration of the generalization capability, Fig.~\ref{fig:exp_results_4} depicts the estimation results of all methods for two evaluation maneuvers of different loading states in- and out-of-distribution. The qualitative comparison in Fig.~\ref{fig:exp_results_4} is underscored by quantitative results for all evaluation maneuvers and for all variables of interest (see Tab.~\ref{tab:results}).\\
In Fig. \ref{fig:exp_results_4}, the plots show measurements and estimates of the articulation angle $\theta$ between truck and semitrailer (top) and the lateral tire force $F_{y_{21}}$ (middle), respectively. 
The bottom plots depict the confidence $\tau_k$ in the \ac{ann} estimates according to \eqref{eq:confidence}.\\
The left plots illustrate the estimation performance in-distribution, as the maneuver is of loading state ``fully loaded''.  
As expected, all estimators perform well. The \ac{ann} estimates are more accurate than the \ac{ekf} estimates, which can be attributed to the high approximation capability of the \ac{ann} and the availability of sufficient training data in the region of the evaluation data. 
This is reflected by a high confidence $\tau$ in the \ac{ann} estimates. Moreover, the \ac{ann} estimation errors correlate with the confidence $\tau$ in the bottom left plot. This is especially visible for the lateral tire force $F_{y_{21}}$ from $129$ to $137\,\mathrm{s}$. For the articulation angle $\theta$, this relationship is less visible, but the \ac{ann} estimation error is generally low.\\
In this scenario, the \ac{hekf} takes advantage of the accurate \ac{ann} estimates because the confidence is generally high. Thus, performance is improved compared to the purely model-based \ac{ekf}. In operation areas with low confidence, however, the \ac{hekf} relies primarily on its model, dismissing potentially inaccurate \ac{ann} ``soft measurements''. 
This behavior can be seen from $129$ to $137\,\mathrm{s}$ in the middle left plot of Fig. \ref{fig:exp_results_4}. 
Initially, the confidence $\tau$ is high, and the \ac{hekf} estimates are similar to the \ac{ann} ``soft measurements''. Thereafter, $\tau$ decreases sharply, causing the \ac{hekf} to rely more on its model-based predictions. 
As the confidence rises again, the impact of the \ac{ann} estimates increases. 
In this way, the \ac{hekf} combines locally good performance of each, \ac{ekf} and \ac{ann}, and has the potential to outperform both.
\begin{table*}[ht]
	\centering
	\parbox{1\textwidth}{
		\caption{\ac{rmse} of the estimates tested with evaluation data in- and out-of-distribution regarding the training data set $\cali{D}_{\ur{tr}}$.}
		\label{tab:results}
		\begin{tabularx}{\textwidth}{p{22mm}p{16mm}p{6mm}p{6mm}p{6mm}p{6mm}p{6mm}p{6mm}p{6mm}p{6mm}p{6mm}p{6mm}p{6mm}p{6mm}}
			\hline\hline
			Evaluation data & Loading state & \multicolumn{3}{l}{\ac{rmse} of $\theta$ in $\ur{rad}$} & \multicolumn{3}{l}{\ac{rmse} of $F_{y_{21}}$ in $\ur{kN}$} & \multicolumn{3}{l}{\ac{rmse} of $F_{y_{23}}$ in $\ur{kN}$} & \multicolumn{3}{l}{\ac{rmse} of $\delta_1$ in $\ur{rad}$} \\ 
			~ & ~ & \ac{ekf} & \ac{ann} & \multicolumn{1}{c|}{H-\ac{ekf}} & \ac{ekf} & \ac{ann} & \multicolumn{1}{c|}{H-\ac{ekf}} & \ac{ekf} & \ac{ann} & \multicolumn{1}{c|}{H-\ac{ekf}} & \ac{ekf} & \ac{ann} & \mbox{H-\ac{ekf}}\\ \hline
			In-distribution & full load & 0.020 & \textbf{0.016} & \textbf{0.016} & 1.693 & 1.570          & \textbf{1.491} & \textbf{1.950} & 2.282          & 2.264          & 0.035 & \textbf{0.031} & \textbf{0.031} \\
            In-distribution & partial load 1 & 0.016 & \textbf{0.010} & 0.011          & 1.753 & \textbf{0.715} & 0.797          & 1.688          & \textbf{1.210} & 1.384          & 0.033 & \textbf{0.026} & \textbf{0.026} \\
            In-distribution & no load & 0.016 & \textbf{0.013} & 0.014          & 1.292 & \textbf{0.594} & 0.825          & 0.835          & \textbf{0.481} & 0.624          & 0.039 & \textbf{0.035} & \textbf{0.035} \\
            Out-of-distribution & partial load 2 & 0.021 & \textbf{0.008} & \textbf{0.008} & 2.095 & 3.478          & \textbf{1.884} & 2.821          & 3.693          & \textbf{2.743} & 0.031 & \textbf{0.026} & 0.027          \\ \hline
            Mean error& ~ & 0.018 & \textbf{0.012} & \textbf{0.012} & 1.708 & 1.589          & \textbf{1.249} & 1.823          & 1.916          & \textbf{1.754} & 0.035 & \textbf{0.030} & \textbf{0.030}\\
			\vspace{-25.5mm}\makecell{Relative mean error}&\multicolumn{13}{r}{\vspace{-1mm}\hspace{5mm}
						\begin{tikzpicture}
						\begin{axis} [ybar,
							bar width=0.85cm,
							enlarge x limits = {abs = .55},
							height=3.5cm,
							width=15.1cm,
							ymax=1,
							ymin=0,
							xticklabels={},
							ytick={0,0.5,1},
							yticklabels={0\,\%,50\,\%,100\,\%},
							axis y line=left,
							hide x axis,
                            ]
							\addplot [line width = .25mm,
							fill = imesblau] 
							coordinates {(0,1) (1,1) (2,0.95149) (3,1) };
							
							\addplot [line width = .25mm,
							fill = imesorange]   
							coordinates {(0,12/18) (1,0.93026) (2,1) (3,0.85223) };
							
							\addplot [line width = .25mm,
							fill = imesgruen]   
							coordinates {(0,12/18) (1,0.7312) (2,0.91509) (3,0.85571)};
							
							\node[above, anchor=west, font=\footnotesize, rotate=90, color=white] at (axis cs:-0.275, 0) {$\frac{0.018}{0.018}$};
							\node[above, anchor=west, font=\footnotesize, rotate=90, color=black] at (axis cs:0, 0) {$\frac{0.012}{0.018}$};
							\node[above, anchor=west, font=\footnotesize, rotate=90, color=black] at (axis cs:0.275, 0) {$\frac{0.012}{0.018}$};
						\end{axis}
				\end{tikzpicture}}\\
			\hline\hline
	\end{tabularx}}
\end{table*}\\
On the right-hand side of Fig. \ref{fig:exp_results_4}, results based on the loading state ``partial load 2'' are presented. This loading state is not included in the training data and differs from ``partial load 1'' regarding the payload distribution (out-of-distribution, see Fig.~\ref{fig:knn_input_space}). 
In this scenario, the \ac{ann} exhibits a significantly lower performance compared to the in-distribution evaluation, which is presumably due to overfitting regarding the input variable $F_{z_2}$. As expected, the large \ac{ann} estimation error is reflected by a generally low confidence $\tau$. 
Comparing the \ac{ann} estimation errors regarding the articulation angle $\theta$ and the lateral tire force $F_{y_{21}}$, the force errors deteriorate more in the out-of-distribution evaluation. This can be attributed to the fact that the kinematic quantity $\theta$ is less reliant on the vertical force $F_{z_2}$ than the dynamic force $F_{y_{21}}$.\\
Considering the \ac{ekf}, the generalization capability out-of-distribution is good. 
The \ac{hekf} inherits this characteristic and disregards ``soft measurements'' with low confidence, resulting in an estimation performance comparable to the purely model-based \ac{ekf}. Thus, the \ac{hekf} takes advantage of accurate (but potentially overfitted) \ac{ann} estimates if evaluated in-distribution and generalizes reliably 
if evaluated out-of-distribution.\\
The previous qualitative results are underpinned quantitatively in Tab.~\ref{tab:results}. The estimation performance of \ac{ekf}, \ac{ann} and \ac{hekf} are compared regarding articulation angle $\theta$, lateral tire forces $F_{y_{21}}$ and $F_{y_{23}}$, as well as steering angle $\delta_1$. All four evaluation maneuvers are considered, including three maneuvers in-distribution regarding the training data $\cali{D}_{\ur{tr}}$ and an additional maneuver out-of-distribution (loading state ``partial load 2'').\\
In-distribution, the \ac{ann} usually outperforms the \ac{ekf}, as expected. Only for the evaluation maneuver with full load, the \ac{ekf} yields better estimation accuracy regarding the lateral tire force $F_{y_{23}}$. This is presumably due to a suboptimally trained \ac{ann} and needs further investigation. The \ac{hekf} takes into account the ``soft measurements'' of the \ac{ann}, resulting in a significantly improved estimation accuracy in-distribution compared to the purely model-based \ac{ekf}. Thus, the \ac{hekf} is at least as good as the traditional methods \ac{ekf} or \ac{ann} and can even outperform both, see for instance, the estimation errors of the lateral tire force $F_{y_{21}}$ based on the evaluation maneuver with full load (see left-hand side in Fig. \ref{fig:exp_results_4} as well).\\
Out-of-distribution, the \ac{ann} estimates of the lateral tire forces $F_{y_{21}}$ and $F_{y_{23}}$ are significantly worse than the \ac{ekf} estimates, indicating overfitting on the training data. Although the \ac{ekf} shows a degradation of estimation accuracy as well, it performs more reliable than the \ac{ann}. The \ac{hekf} inherits this characteristic due to low confidence levels in the \ac{ann} estimates in the out-of-distribution scenario. 
For the articulation angle $\theta$ and the steering angle $\delta_1$, which are kinematic quantities, the \ac{ann} shows good performance even if evaluated out-of-distribution, which is presumably due to a low impact of the vertical force input $F_{z_2}$ on the kinematic quantities.\\ 
Thus, the experimental comparison in Tab. \ref{tab:results} suggests that  
the \ac{ann} allows for accurate approximation of the semitrailer behavior in-distribution regarding the training data set $\cali{D}_{\ur{tr}}$, but suffers from overfitting. In contrast, the \ac{ekf} provides better generalization capability if evaluated out-of-distribution at the cost of less estimation accuracy in-distribution.\\ 
The \ac{hekf} alleviates the limitations of both methods and allows for accurate estimation in-distribution while reliably generalizing out-of-distribution. Regarding kinematic quantities in the present truck-semitrailer application, the \ac{hekf} exhibits comparable performance as the \ac{ann} estimation. 
Considering the dynamic lateral tire forces $F_{y_{21}}$ and $F_{y_{23}}$, the \ac{hekf} clearly outperforms both comparison methods, \ac{ekf} and \ac{ann}, in the experimental analysis. 
\section{Conclusion and Outlook} \label{sec:conclusion} 
In real-world systems, it is difficult to obtain training data that spans the whole operation region of a system. Thus, overfitting is likely to occur in learning-based estimation, resulting in bad generalization capability out-of-distribution. In contrast, the estimation performance of purely model-based methods is usually limited by the model accuracy.\\
In this paper, we analyze model-based, learning-based and hybrid estimation schemes regarding their accuracy and generalization capability in- and out-of-distribution. 
Beyond the state of research, the comparison study is conducted using extensive experimental data of a full-scale truck-semitrailer combination with various loading states.
To enable reliable estimation out-of-distribution, the confidence-based hybrid estimation framework in \cite{Sieberg.2022, Sieberg.2022b} is employed and extended. In particular, a new \ac{knn}-based scheme is proposed to quantify the confidence in the ``soft measurements'' from an \ac{ann} estimator and the scheme is enhanced for consideration of multidimensional \ac{ann} estimates.\\
The experimental results show that the \ac{ann} performs best if evaluated in-distribution but generalizes badly to other data, indicating overfitting. The purely model-based \ac{ekf} provides reliable estimates out-of-distribution, but has a higher estimation error level in-distribution. 
In comparison, the proposed \ac{hekf} improves estimation performance in-distribution and provides reliable estimates (generalization) out-of-distribution regarding the training data, suggesting that the proposed \ac{hekf} combines the advantages of the considered model-based and learning-based estimation schemes.\\
By enabling reliable estimation out-of-distribution, the \ac{hekf} has the potential to facilitate development of \ac{adas}, autonomous driving functions, and safety features in the current application to truck-semitrailer combinations and beyond. 
Future work may focus on investigating the computational complexity of the \ac{hekf}. Besides, other \ac{ann} structures may be tailored for estimation and examined regarding their generalization capability. A particularly interesting choice are physics-informed neural networks which allow incorporating model knowledge and have shown promising learning capability in physical systems with limited real-world training data \cite{NICODEMUS2022331}. 

\section*{Acknowledgment}
This work is part of the project IdenT (19$|$19008A, 19$|$19008B), which is funded by the German Federal Ministry for Economic Affairs and Climate Action based on a resolution of the German Bundestag.

\balance

\bibliographystyle{IEEEtran}
\bibliography{IEEEabrv,references_ECC}

\end{document}